\documentstyle[preprint,epsfig,epsf,aps,floats]{revtex}
\newcommand{\be}{\begin{equation}}
\newcommand{\ee}{\end{equation}}
\newcommand{\gsim}{\, \raisebox{-0.8ex}{$\stackrel{\textstyle >}{\sim}$ }}
\newcommand{\lsim}{\, \, \raisebox{-0.8ex}{$\stackrel{\textstyle <}{\sim}$ }}

\newcommand{\roughly}[1]%
{\mathrel{\raise.4ex\hbox{$#1$\kern-.75em\lower1ex\hbox{$\sim$}}}} 

\newcommand\CL{{\cal L}}

\newcommand\beq{\begin{eqnarray}} 
\newcommand\eeq{\end{eqnarray}} 
\newcommand\eqn[1]{\label{eq:#1}}

\def\Dsl{\,\raise.15ex \hbox{/}\mkern-12.8mu D} 
\newcommand\Tr{{\rm Tr\,}}

\def\fm3{fm$^{-3}$}

\def\ni{\noindent}
\begin{document}
%
\preprint{\vbox{\hbox{MIT-CTP-3252}}}

\bigskip
\bigskip

\title{Neutrino Rates in Color Flavor Locked Quark Matter }
\author{Sanjay Reddy$^1$, Mariusz Sadzikowski$^{1,2}$
and 
Motoi Tachibana$^{1,3}$}

\address{$^1$Center for Theoretical Physics, Massachusetts 
Institute of Technology, Cambridge, MA 02139 \\ $^2$Institute of
Nuclear Physics, Radzikowskiego, 152, 31-342, Krak\'ow, Poland. \\
$^3$ Yukawa Institute for Theoretical Physics, Kyoto University, Kyoto
606-8502, Japan. }
\maketitle

\begin{abstract} 
We study weak interaction rates involving Goldstone bosons in Color
Flavor Locked (CFL) quark matter. Neutrino mean free path and the rate
of energy loss due to neutrino emission in a thermal plasma of CFL
pions and kaons is calculated. We find that in addition to neutrino
scattering off thermal mesons, novel Cherenkov like processes wherein
mesons are either emitted or absorbed contribute to the neutrino
opacity. Lack of Lorentz invariance in the medium and loss of
rotational invariance for processes involving mesons moving relative
to the medium allow for novel processes such as $\pi^0 \rightarrow \nu
\bar{\nu}$ and $e^- \pi^+ \rightarrow \nu_e$.  We explore and comment
on various astrophysical implications of our finding.
\end{abstract}


\section{Introduction}
QCD at high baryon density is expected to be a color superconductor.
For three massless flavors, a symmetric ground state called the Color
Flavor Locked (CFL) phase, in which BCS like pairing involves all nine
quarks, is favored \cite{Alford:1999mk}. This, color superconducting,
phase is characterized by a gap in the quark excitation
spectrum. Model calculations indicate that the gap $\Delta \sim 100$
MeV for a quark chemical potential $\mu \sim 500$ MeV
\cite{Alford:1997zt,Rapp:1997zu}.  In this phase the ${ SU(3)_{\rm
color}} \times SU(3)_L \times SU(3)_R \times U(1)_B$ symmetry of QCD
is broken down to the global diagonal $SU(3)$ symmetry due to pairing
between quarks at the Fermi surface. Gluons become massive by the
Higgs mechanism. The lightest excitations in this phase are the nonet
of pseudo-Goldstone bosons transforming under the unbroken, global
diagonal $SU(3)$ as an octet plus a singlet and a massless mode
associated with the breaking of the global $U(1)_B$ symmetry (For a
recent review see \cite{Rajagopal:2000wf}).

The lack of quark particle-hole excitations in dense quark matter at
temperatures less than the critical temperature for color
superconductivity ($T_c \sim 0.6 \Delta$) will affect thermodynamic
and transport properties of this phase. This can have important
astrophysical implications if quark matter were to exist in the core
of a neutron star.  Several authors have recently explored some
aspects of how color superconductivity might impact neutron star
observables.  These works have provided insight on: role of color
superconductivity on the phase transition density; the nature of the
interface between nuclear matter and CFL matter \cite{Alford:2001zr};
its response to magnetic fields \cite{Alford:1999pb}; and the thermal
evolution of both young and old neutron
stars\cite{Blaschke:1999qx,Page:2000wt,Carter:2000xf}. In this article
we calculate weak interaction rates for neutrino production and
scattering in the CFL phase and contrast it with earlier estimates of
similar rates in normal(unpaired) quark matter.

Neutrinos play a central role in the early and late time thermal
evolution of neutron stars. Weak interaction rates in the
superconducting phase are therefore essential to make connections
between color superconductivity and observable aspects associated with
neutron star thermal evolution.  Neutron stars are born in the
aftermath of a core collapse supernova explosion. The inner core of a
newly born neutron star is characterized by a temperature ${\rm T}\sim
30$ MeV and a lepton fraction (lepton number/baryon number) ${\rm Y_L}
\sim 0.3$ implying $\mu_e \equiv \mu_{\nu_e}-\mu_Q \sim \mu_{\nu_e}
\sim 200$~MeV. The high temperature and finite lepton chemical
potentials are a consequence of neutrino trapping during gravitational
collapse. The ensuing thermal evolution of the newly born neutron
star, during which it emits neutrinos copiously, has generated much
recent interest \cite{Burrows:1986me,Pons:2001ar}. Several aspects of
this early evolution can be probed directly since neutrinos emitted
during the first several tens of seconds can be detected in
terrestrial detectors such as Super-Kamiokande and SNO.  This study is
motivated by the prospect that were Color-Flavor-Locked quark matter
to exist in neutron stars at early times it would result in observable
and discernible effects on the supernova neutrino emission.

We can expect significant differences in the weak interaction rates
between the normal and the CFL phases of quark matter since the latter is
characterized by a large gap in the quark excitation spectrum. Thus,
unlike in the normal phase where quark excitations near the Fermi
surface provide the dominant contribution to the weak interaction
rates, in the CFL phase, it is the dynamics of the low energy
collective states--- the flavor pseudo-Goldstone bosons that are
relevant. As a first step towards understanding the thermal and
transport properties of this phase which are of relevance to core collapse
supernova studies, we identify and calculate the weak interaction
rates for neutrino production and scattering. The article is organized
as follows: In \S2 we briefly describe the effective theory describing
Goldstone bosons in the CFL phase; In \S3 we calculate the rate of
neutrino reactions of interest; and in \S4 we discuss how our findings
might affect the early evolution of a newly born neutron star.

\section{Effective Theory For Goldstone Bosons}
There are several articles that describe in detail the effective
theory for the Goldstone bosons in Color-Flavor-Locked quark matter
\cite{Hong:1999dk,Casalbuoni:1999wu,Rho:1999xf,Son:2000cm,Manuel:2000wm,Beane:2000ms,Schafer:2001za}.
We briefly review some aspects of the effective theory in this
section.  It is possible to parameterize low energy excitations about
the $SU(3)$ symmetric CFL ground state in terms of the two fields
$B=H / (\sqrt{24}f_H)$ and $\Sigma=e^{2i({\pi}/f_\pi+\eta'/f_A)}$,
representing the Goldstone bosons of broken baryon number $H$, and
of broken chiral symmetry, the pseudo-scalar octet ${ \pi}$, and the
pseudo-Goldstone boson $\eta'$, arising from broken approximate
$U(1)_A$ symmetry. Turning on nonzero quark masses, explicitly breaks
chiral symmetry and induces a gap in the spectrum as the Goldstone
bosons acquire a mass due to this explicit breaking of chiral
symmetry. In addition, dissimilar quark masses induce new stresses on
the system, acting in a manner analogous to an applied flavor chemical
potential and favoring meson condensation. This was recently pointed
out by Bedaque and Schafer \cite{Bedaque:2001je}. In this work we
include the stress induced by the strange quark mass, but we will
assume that it is not strong enough to result in the meson
condensation
\footnote{If $K^0$ condensation occurs, as discussed in 
Ref.\cite{Bedaque:2001je,Kaplan:2001qk} the ground state is
reorganized and the excitation spectrum is modified. The weak
interaction rates in this phase is currently under investigation and
will be reported elsewhere.}.

The leading terms of the effective Lagrangian describing the octet
Goldstone boson field $\pi$ is given by 

\beq \eqn{leff} \CL &=&
\frac{1}{4}f_{\pi}^2 \left[\Tr \nabla_0 \Sigma \nabla_0 \Sigma^\dagger
- v^2 \Tr \vec{\nabla}\Sigma \cdot \vec{\nabla} \Sigma^\dagger \right]
\ \, \nonumber \\ && + f_{\pi}^2 \left[\frac{a}{2} \Tr \tilde{M}
\left(\Sigma + \Sigma^\dagger\right) + \frac{\chi}{2} \Tr {M}
\left(\Sigma +\Sigma^\dagger\right)\right] \ \, \nonumber \\ & &
\nabla_0\Sigma = \partial_0 \Sigma - i\left[ X_L \Sigma - \Sigma X_R
\right] \,.  
\eeq 

The decay constant $f_\pi$ has been computed previously
\cite{Son:2000cm} and is proportional to the quark chemical
potential. $X_{L,R}$ are the Bedaque-Schafer terms: $X_L = -\frac{M
M^\dagger}{2\mu}\ ,\quad X_R = -\frac{ M^\dagger M}{2\mu}\ \,, $ and
$\tilde{M}= |M|M^{-1}$. A finite baryon chemical potential breaks
Lorentz invariance of the effective theory. The temporal and spatial
decay constants can thereby differ. This difference is encoded in the
velocity factor $v$ being different from unity. An explicit
calculation shows that $v=1/\sqrt{3}$ and is common to all Goldstone
bosons, including the massless $U(1)_B$ Goldstone
boson\cite{Son:2000cm}.

At asymptotic densities, where the instanton induced interactions are
highly suppressed and the $U(1)_A$ symmetry is restored, the leading
contributions to meson masses arise from the Tr$ \tilde{M} \Sigma$
operator whose coefficient $a$ has been computed and is given by
$a=3\frac{\Delta^2}{\pi^2 f_{\pi}^2}$
\cite{Son:2000cm,Schafer:2001za}. At densities of relevance to neutron
stars the instanton interaction may become relevant. In this case a
$<\bar{q} q>$ condensate is induced
\cite{Manuel:2000wm,Schafer:2002ty} and consequently the meson mass
term can receive a contribution from the operator Tr$ M \Sigma$.  Its
coefficient $\chi$ at low density is sensitive to the instanton size
distribution and form factors. Current estimates indicate that the
instanton contribution to the $K^0$ mass lies in the range $5-120$ MeV
\cite{Schafer:2002ty}. The meson masses are explicitly given by

\beq
m^2_{\pi^\pm} &=& a (m_u + m_d)m_s + \chi (m_u+m_d) \cr m^2_{K^\pm}
&=& a (m_u + m_s)m_d + \chi (m_u+m_s) \cr m^2_{K^0} &=& a (m_d +
m_s)m_u + \chi (m_d+m_s)\,.  
\eqn{masses}
\eeq

In this article we will assume the instanton contribution to the $K^0$
mass is $\sim 50$ MeV (corresponding to $\chi \sim 15 $ MeV) at
$\mu=400$ MeV and $\Delta=100$ MeV.  With this choice, the kaon mass
is too large to allow for $K^0$ condensation.

To incorporate weak interactions, we gauge the Chiral Lagrangian
in the usual way by replacing the covariant derivative by
\cite{Casalbuoni:2000jn}

\begin{eqnarray}
D_\mu \Sigma = && \nabla_\mu \Sigma - \frac{ig}{\sqrt{2}}
(W^{+}_\mu\tau^+ + W^{-}_\mu\tau^-)\Sigma \nonumber \\
&&- \frac{ig}{\cos{\theta_W}} Z_\mu
(\tau_3 \Sigma - \sin^2{\theta_W}~ [Q,\Sigma])
- i~\tilde{e} \tilde{A}~[Q,\Sigma] 
\end{eqnarray}

The time component of $\nabla_\mu $ includes the Bedaque-Schafer
term as described above. The fields $W^{\pm}_\mu ,Z_\mu $ describe
weak gauge bosons. The charge matrix is diagonal $Q =
\mbox{diag}(\frac{2}{3},-\frac{1}{3},-\frac{1}{3})$ as well as
weak-isospin matrix $\tau_3 =\frac{1}{2}\mbox{diag}(1,-1,-1)$ whereas
$\tau^+$ and $\tau^-$ are the isospin raising and lowering operators
which incorporate Cabbibo mixing. The weak coupling constant is related
to Fermi coupling constant via the standard relation $\sqrt{2} g^2=8
G_F M_{W}^2$ where $M_W$ is a mass of the $W$ gauge boson and the mass
of $Z$ boson $M_Z \cos{\theta_W} = M_W$ where $\theta_W$ is Weinberg
angle. The last term is the modified electromagnetic coupling of
mesons to the massless $\tilde{A}$ photon in the CFL phase, where the
charge $\tilde{e}=e \cos{\theta}$ and $\theta$ is the mixing angle
between the original photon and the eighth gluon \cite{Alford:1999pb}.

For momenta small compared to $f_\pi$ we can expand the nonlinear
chiral Lagrangian to classify diagrams as the first order
(proportional to $f_\pi $) and the second order (independent of $f_\pi
$).  The first order diagrams, involving two leptons and a meson are
shown in the Fig. 1. In Fig. 1(a) the charged current decay of charged
pions and kaons is shown. Fig. 1(b) shows the neutral current decay of
$\pi^0 \rightarrow \nu \bar{\nu}$.  In vacuum, the latter
process is forbidden by angular momentum conservation. However, as we
will show later, it is allowed for finite momentum pions in the CFL
medium. Diagrams in the Fig. 1(c) and 1(d) show processes involving
two mesons.  The process in Fig. 1(c) is the two body correction to
the charged current decay of the decay of charged kaons and pions
interacting with the neutral mesons. And Fig. 1(d) depicts neutrino pair 
emission from the annihilation of charged Goldstone bosons 
$K^\pm$ and $\pi^\pm $ and neutrino-meson scattering. 

\begin{figure}[t]
\begin{center}
\includegraphics[width=.8\textwidth,height=0.25\textheight]{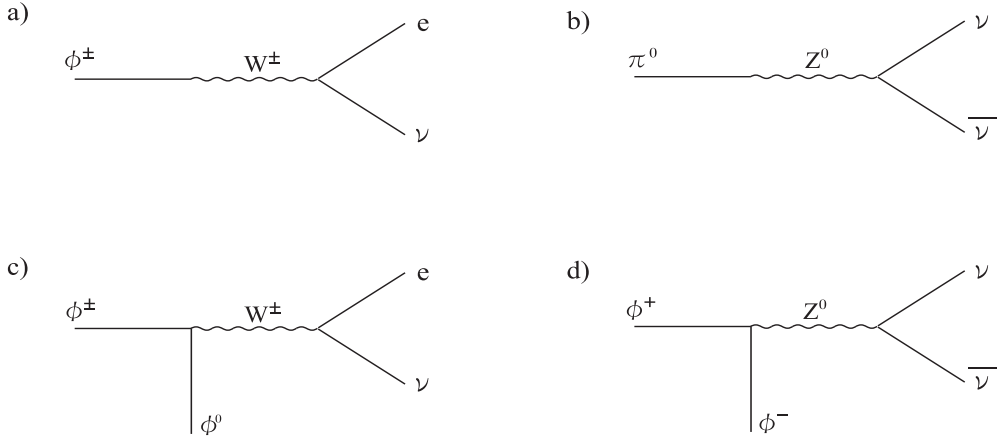}
\end{center}
\caption{Feynman graphs showing the coupling of the neutral and charged 
leptonic current to one and two meson states.}
\label{feyn}
\end{figure}

The amplitudes for the leading order processes are given by 

\beq
A_{\pi^0\rightarrow \nu \bar{\nu} } &=&
~\frac{G_F}{\sqrt{2}}~f_\pi~\tilde{p}_\mu ~j_Z^\mu \\ \nonumber
  A_{\pi^\pm\rightarrow e\nu } &=& ~G_F~f_\pi~ \cos{\theta_C} ~\tilde{p}_\mu
  ~j_W^\mu \\\nonumber A_{K^\pm\rightarrow e\nu } &=&~ G_F ~f_\pi~
  \sin{\theta_C }~p_\mu ~j_W^\mu 
\eeq 

where $\tilde{p}_\mu=(E,v^2\vec{p})$ is the modified four-momentum of
Goldstone boson and $j_W^\mu$ and $j_Z^\mu$ describe the charged
leptonic current and neutral leptonic currents.  $\theta_C$ is the
Cabbibo mixing angle. Note that the meson "four momenta" that appear
in the matrix element do not correspond to the on shell four momenta
of the mesons. This is because the covariant derivatives contains the
in medium velocity and for the case of kaons, the energy shift arising
from the Bedaque-Schafer term.

At next to leading order, the amplitude is independent of $f_\pi$. The amplitude for the charged current process is given by
\begin{equation}
A_{\Phi^\pm\Phi^0\rightarrow e\nu } = ~-iC~\frac{G_F}{2}~ 
(\tilde{p}_1-\tilde{p}_2)_\mu~ j_W^\mu
\end{equation}
where the coupling coefficient $C=\sin{\theta_C }$ for $\pi^0K^-
,\pi^0K^+$; $C=\sqrt{2}\sin{\theta_C }$ for $\pi^-\bar{K}^0,
\pi^+K^0$; $C=\sqrt{2}\cos{\theta_C }$ for $K^-K^0, K^+\bar{K}^0$ and
$C=2\cos{\theta_C }$ for $\pi^-\pi^0,\pi^+\pi^0$. Neutrinos couple to
the charged mesons via the neutral current.  This leads to process
such as the annihilation of $\pi^+\pi^- \rightarrow
\nu \bar{\nu}$ and $K^+ K^- \rightarrow \nu \bar{\nu}$, whose amplitude is 
given by
\begin{equation}
A_{\Phi^+\Phi^- \rightarrow \nu\bar{\nu } } = ~-i\frac{G_F}{\sqrt{2}}
~(1-2\sin^2{\theta_W})~  (\tilde{p}_1-\tilde{p}_2)_\mu~ j_Z^\mu
\end{equation}
where $p_1,p_2$ are momenta of Goldstone bosons. It also gives rise to 
neutral current neutrino-meson scattering given by the amplitude
\begin{equation}
A_{\nu \Phi^{\pm} \rightarrow \nu \Phi^{\pm} } = ~-i\frac{G_F}{\sqrt{2}}
~(1-2\sin^2{\theta_W})~  (\tilde{p}_1+\tilde{p}_2)_\mu~ j_Z^\mu \,.
\end{equation}

In addition to the flavor octet of Goldstone bosons, the massless
Goldstone boson associated with spontaneous breaking of $U(1)_B$
couples to the weak neutral current. This is because the weak isospin
current contains a flavor singlet component. Amplitude for processes 
involving the $U(1)_B$ Goldstone boson $H$ and the neutrino neutral 
current is given by

\be 
A_{H \nu \bar{\nu}}=
~\frac{4}{\sqrt{3}}~G_F~f_H ~ \tilde{p}_{\mu}~j_Z^{\mu}\,, 
\ee 

\ni where $\tilde{p}_{\mu}=(E,v^2 \vec{p})$ is the modified four
momentum of the Goldstone boson. The decay constant for the $U(1)_B$
Goldstone boson has also been computed in earlier work and is given by
$f^2_H=3 \mu^2/(8 \pi^2)$\cite{Son:2000cm}. Further, we note that a $H
H Z^0$ vertex is absent in the lowest order Lagrangian ($H$ boson does
not carry baryon number, isospin or hypercharge).

\section{Thermodynamics and Neutrino Rates}
The dispersion relations for Goldstone modes in the CFL phase are
unusual.  They are easily computed by expanding the Lagrangian to
second order in meson fields and are given by 
\beq
E_{K^{\pm}}&=& \mp \frac{ m_s^2}{2\mu} + \sqrt{v^2p^2 + m_{K^\pm}^2} \,,~~~~~
E_{\pi^{\pm}}=\sqrt{v^2p^2 + m_{\pi^\pm}^2} \,,\nonumber \\
E_{K^{0}}&=&-\frac{m_s^2}{2\mu} + \sqrt{v^2p^2 + m_{K^0}^2} ~~~~~\,,
E_{\bar{K^{0}}}=\frac{m_s^2}{2\mu} + \sqrt{v^2p^2 + m_{K^0}^2} \,.
\eeq
They violate Lorentz invariance and the induced effective chemical
potential arising from the analysis of Bedaque and Schafer 
\cite{Bedaque:2001je} breaks the energy degeneracy of the positive and
negative charged kaons. This making the $K^+$ lighter than the $K^-$ and 
naturally results in an excess positive charge in the
meson sector at finite temperature. Electric charge neutrality of this
phase demands electrons and consequently an electric charge chemical
potential is induced. This novel phenomena, akin to the thermoelectric
effect, modifies the number densities of the individual mesons in the
plasma.

Kaon, pion and the total electric charge density of the meson gas,
normalized to the photon number are shown in Fig. 2. The results are
shown as a function of temperature for a quark chemical potential of
400 MeV (corresponding to $f_\pi= 83 $ MeV) at which we have chosen
the pairing gap $\Delta=100$ MeV. The quark masses are set at
$m_u=3.75$ MeV $m_d=7.5$ MeV and $m_s=150$ MeV and instanton induced
coefficient of the Tr$M\Sigma$ operator $\chi=16$ MeV as mentioned
earlier. With these assumptions, at zero temperature the rest energy
of mesons are given by:$E_{\pi^{\pm}}\simeq 30$ MeV; $E_{K^+}\simeq
26$ MeV; $E_{K^-}\simeq 82 $ MeV ; $ E_{K^0}\simeq 24$ MeV and
$E_{\bar{K^0}}\simeq 80$ MeV. Consequently, the kaon number is
dominated by the $K^0$ and $K^+$ mesons and a finite electron chemical
potential favors $\pi^-$ mesons over the $\pi^+$. It is surprising
that, with increasing temperature, the plasma has more mesons than
photons. Despite being massive, the meson number is enhanced for two
reasons. First, the velocity factor $v=1/\sqrt{3}$ results in a soft
dispersion relation resulting in reduced Boltzmann suppression of the
high momentum modes. The number density is enhanced by a factor
$1/v^3$. For the same reason, the density of massless $H$ bosons (not
shown in the figure) is also larger than the photon density by the factor
$1/(2 v^3)$.  Secondly, the Bedaque-Schafer term which acts like an
effective chemical potential favoring strange quarks, and the induced
negative electric charge chemical potential required to maintain
charge neutrality enhances the number of mesons with strange quarks
and negative charge.

\begin{figure}[t]
\begin{center}
\includegraphics[width=.8\textwidth]{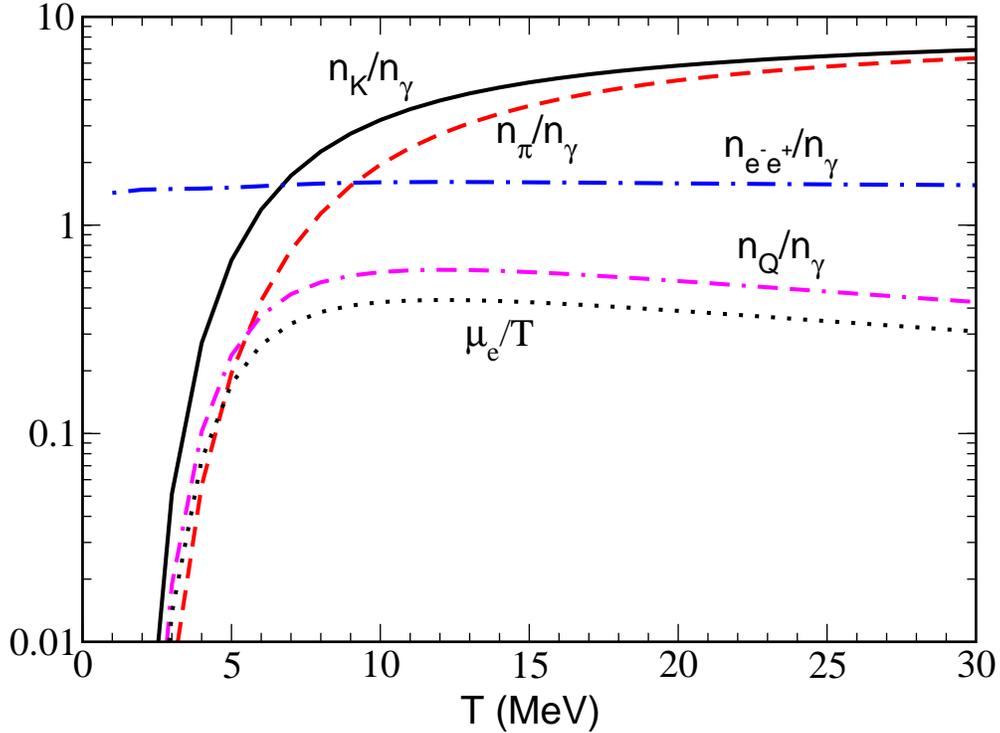}
\end{center}
\caption{Number densities of the mesons, the charge density and the number of 
$e^+ e^-$ pairs, normalized by the photon number density. The electron
chemical potential required to maintain electric charge neutrality at
finite temperature is also shown.}
\label{number}
\end{figure}

Prior to discussing neutrino rates in the finite temperature
plasma we briefly comment on the temperature range over which our
analysis is expected to be valid. The effective theory description of
the finite temperature phase is valid only if the typical meson energy
is small compared to $2 \Delta$. For larger energies, meson
propagation is strongly attenuated since they can decay into quark
quasi-particle excitations. Further, with increasing temperature, we
should expect coefficients of the effective theory such as $
f_\pi, ~v$ and the meson masses to change. On general grounds, we expect
these changes to become relevant as $T$ approaches $T_c$. In our
analysis, we ignore these changes and restrict ourselves to
temperatures $T$ small compared to $T_c$. At $\mu=400$ MeV and $\Delta =
100 $ MeV corresponding to $T_c \sim 60$ MeV we expect our analysis to
provide a fair description of the plasma for $ T \lsim 30$ MeV
\footnote{Assuming $\Delta(T) = \Delta_0\sqrt{1-(T/T_c)^2}$, at $T=30$
MeV the gap is reduced by only 15 percent.}.

\subsection{Neutrino Opacity}
Novel processes such as the $\nu \rightarrow H \nu $, $\nu
\rightarrow \pi^0 \nu$ and $ \nu_e \rightarrow \pi^+ e^-$ are allowed
in this phase owing to the fact that mesons can have a space like
dispersion relation. These processes can be thought of as arising due
to Cherenkov radiation of mesons. Dispersion relations for the
pions and kaons indicate that they are space like four momenta when
their 3-momentum satisfy: 

\beq 
p_\pi > \frac{m_\pi}{\sqrt{1-v^2}}
~~~;~~~ p_K > \frac{\sqrt{m_K^2(1-v^2) + X^2} - X}{1-v^2} \,,
\label{space}
\eeq 

\ni where $X=m_s^2/(2\mu)$ for ($K^+,K^0$) and $X=-m_s^2/(2\mu)$ for
($K^-,\bar{K}^0$).  Consequently, only high energy neutrinos, with
energy greater than $E_{TH} = m_\pi/\sqrt{1-v^2}$, can emit pions as
Cherenkov radiation.  In contrast, the massless $U(1)_B$ Goldstone
boson has a space like dispersion relation for all momenta. Thus,
neutrinos of all energies can Cherenkov radiate $H$ bosons as they
traverse the dense CFL medium.

We can define the neutrino mean free path for these processes as the neutrino
velocity times the rate of emission of Cherenkov mesons. This is given by 

\be
\label{cherenkov}
\frac{1}{\lambda_{\nu \rightarrow \phi l}(E_\nu)} = \frac{1}{2 E_\nu}\int
\frac{d^3p_\phi}{(2\pi)^3 2E_\phi}\int\frac{d^3p_l}{(2\pi)^3 2E_l}
(2\pi)^4~\delta^4(P_\nu - P_\phi -P_l) |A_{\nu\rightarrow \phi l}|^2 
\ee 

\ni where $E_\nu$ is the initial neutrino energy, $P_\phi$ is the meson
four momentum and $P_l$ is the final state lepton four momentum. We
label the final state lepton as $l$ to account for the fact that it
could be either a neutrino or an electron. The amplitude for this
process was calculated earlier and is proportional to $\mu$. Since
these processes do not have mesons in the initial state they can
occur at zero temperature. The rate for the process
depends on the neutrino energy and is independent of the temperature
in so far as the meson masses and $f_\pi$ is independent of
temperature. Kinematics and 1 relatively strong $H \nu \bar{\nu}$
coupling (approximately thrice as a large as the $\pi^0 \nu \bar{\nu}$
coupling) makes Cherenkov radiation of the massless $H$ bosons the 
dominant reaction of this type. Neutrino mean free
path due to the reaction $\nu \rightarrow H \nu$ can be calculated
analytically using Eq. (\ref{cherenkov}) and is given by

\be \frac{1}{\lambda_{\nu\rightarrow H \nu}(E_\nu)}=
~\frac{256}{45\pi}~\left[\frac{v(1-v)^2(1+\frac{v}{4})}{(1+v)^2}
\right]~G_F^2~f_H^2
~~E_{\nu}^3 
\ee

Neutrinos of all energies can absorb a thermal meson and scatter into
either a final state neutrino by neutral current processes like $\nu +
H \rightarrow \nu$ and $\nu+ \pi^0 \rightarrow \nu$ or via the
charged current reaction into a final state electron by the process
$\nu_e + \pi^- \rightarrow e^-$.  These processes are temperature
dependent as they are proportional to the density of mesons in the
initial state. Mean free path due to these processes, which we
collectively refer to as Cherenkov absorption, can be computed using

\be 
\frac{1}{\lambda_{\nu\phi\rightarrow l}(E_\nu)} = \frac{1}{2 E_\nu}\int
\frac{d^3p_\phi}{(2\pi)^3 2E_\phi}~f(E_\phi)\int\frac{d^3p_l}{(2\pi)^3 2E_l}
(2\pi)^4~\delta^4(P_\nu+P_\phi-P_l) |A_{\nu \phi\rightarrow l}|^2\nonumber 
\ee

\ni where $f(E_\phi)$ is the Bose distribution function for the initial
state mesons. Reactions involving the $H$ boson dominate over
other Cherenkov absorption processes due to their larger population
and stronger coupling to the neutral current. For this case, we find 
the neutrino mean free paths is given by 

\be
\frac{1}{\lambda_{\nu H \rightarrow \nu}(E_\nu)}=
\frac{128}{3\pi}\left[\frac{v~(1+v)^2}{(1-v)}\right]
\left[g_2(\gamma)+
\frac{2v}{(1-v)}g_3(\gamma)-\frac{(1+v)}{(1-v)}g_4(\gamma)\right]  
~G_F^2~f_H^2~E_\nu^3 \,,
\ee

\ni where $\gamma=2vE_\nu/(1-v)T$ and the integrals $g_n(\gamma)$ are defined
by the relation

\be
g_n(\gamma) = \int_{0}^{1} dx ~\frac{x^n}{\exp{(\gamma x)}-1} \,. 
\ee

In the limiting cases of high ($E_\nu \gg T$) and low ($ E_\nu \ll T$)
neutrino energy the above integral can be performed analytically. For
$E_\nu \gg T$ we find that

\be 
\frac{1}{\lambda_{\nu H \rightarrow \nu}(E_\nu)} =
\frac{32 \zeta(3)}{3\pi}\left[\frac{(1-v^2)^2}{v^2}\right]
\left[1 + 
\frac{\pi^4}{30 \zeta(3)}~\frac{T}{E_\nu} - 
\frac{3\zeta(5)}{\zeta(3)}~\frac{(1-v^2)}{v^2}
~\frac{T^2}{E_\nu^2}\right]~G_F^2~f_H^2~T^3 \,, 
\ee 

\ni and for the case when $E_\nu \ll T$ we find 

\be 
\frac{1}{\lambda_{\nu H \rightarrow \nu}(E_\nu)}=\frac{16}{3\pi}
\frac{(1+v)^2(1-\frac{v}{3}) }{(1-v)}~G_F^2~f_H^2~T~E_\nu^2 \,.  
\ee

In contrast to processes involving the emission or absorption of
mesons by neutrinos, the usual scattering process involves the
coupling of the neutrino current to two mesons. As noted earlier, the
amplitude for these processes vanishes for the $H$ meson and is
suppressed by the factor $p/f_\pi$ where $p$ is the meson momentum for
the flavor octet mesons. In the later case, despite the suppression,
the reaction could be important because the kinematics is not
restricted to space like mesons. Neutrino mean free path for the
scattering process is given by

\beq \frac{1}{\lambda_ {\nu
    \phi^{\pm}\rightarrow \nu \phi^{\pm} }(E_\nu)} &=& \frac{1}{2 E_\nu}\int
\frac{d^3p_\phi}{(2\pi)^3 2E_\phi}~f(E_\phi)\int \frac{d^3p'_\nu}{(2\pi)^3
  2E'_\nu}
\int \frac{d^3p'_\phi}{(2\pi)^3 2E'_\phi} \nonumber \\
&\times&(2\pi)^4~\delta^4(P_\nu+P_\phi-P'_\phi-P_\nu) |A_{\nu
  \phi^{\pm}\rightarrow \nu \phi^{\pm}}|^2 
\eeq 

\begin{figure}[t]
\begin{center}
\includegraphics[width=.8\textwidth]{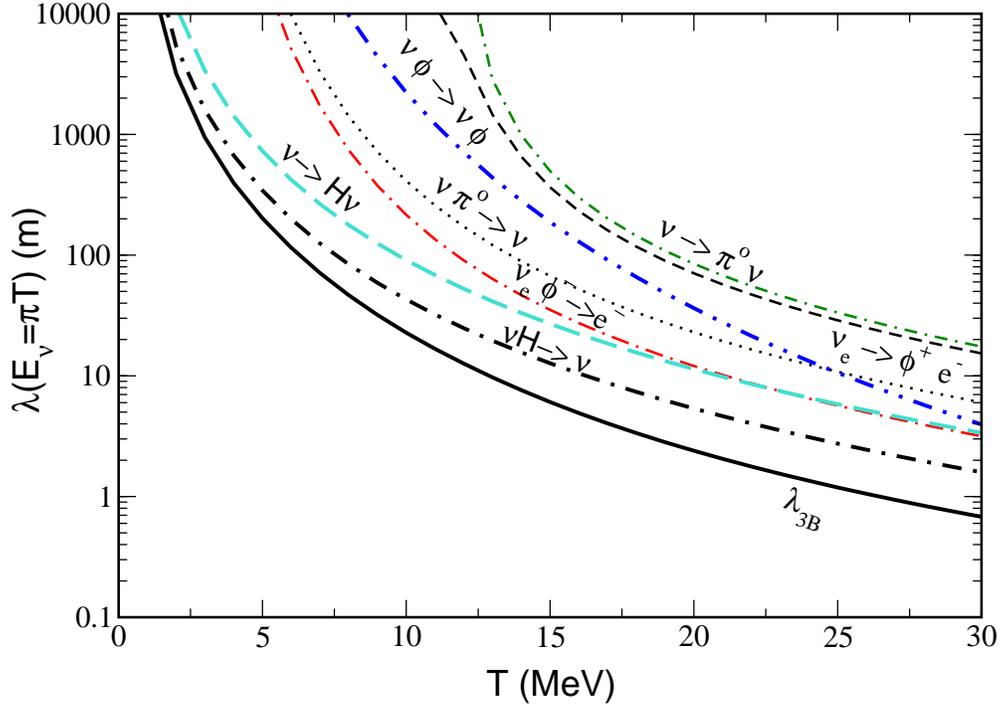}
\end{center}
\caption{Neutrino mean free path in a CFL meson plasma as a function
of temperature. The neutrino energy $E_\nu=\pi T$ and is characteristic of 
a thermal neutrino.}
\label{path}
\end{figure}

The resulting neutrino mean free path for all three classes of
processes discussed above is shown in Fig. 3. The results are
presented as a function of the ambient temperature. We assume that the
typical neutrino energy is determined by the local temperature and set
$E_\nu= \pi T$ since this corresponds to mean energy of neutrinos in
thermal equilibrium. Note, the mean free path due to Cherenkov
radiation of mesons has no intrinsic temperature
dependence. Temperature dependence of the result shown in Fig. 3
arises because we have set $E_\nu= \pi T$. At low temperature, we
expect Cherenkov reactions involving the massless Goldstone boson to
dominate.  Interestingly, however, we find that this is the case even
for $T\sim 30$ MeV. The total contribution to opacity from novel
Cherenkov processes is also shown in the figure as a solid curve
labeled $\lambda_{3B}$. The Cherenkov absorption reaction, $\nu H
\rightarrow \nu $, makes the largest contribution to the neutrino
opacity over the temperature range $T=1-30$ MeV. Reactions involving
kaons and pions become comparable only for $T\gsim 15$ MeV.  Further,
charged current reactions involving the absorption or emission of
kaons are Cabbibo suppressed and contribute to less than a few percent
of the total rate. Scattering reactions are found to be negligible for
$T \lsim 10$ MeV and make a 20\% contribution to the total opacity at
$T\sim 30$ MeV.

\subsection{Neutrino Emissivity}
Charged current decays of pions and kaons, and the novel neutral
current decay of $\pi^0$ are the leading one body process contributing
to neutrino emission. In vacuum, the amplitude for similar processes
are proportional to the lepton mass due to angular momentum
conservation. However, as discussed earlier, the dispersion relations
for Goldstone modes violate Lorentz invariance. Consequently we find
that the decay of finite momentum pions and kaons is not suppressed by
the electron mass (note that decay into muons is highly suppressed
because the meson mass and the temperature are less than the mass of
the muon). This can be understood by noting that Goldstone modes at
finite density are collective excitations associated with deformations
of the Fermi surface. In the rest frame of a meson that has finite
momentum relative to the medium the Fermi Surface is not spherically
symmetric. This breaks rotational invariance and angular momentum is
no longer a good quantum number to describe the meson state. As a
result, decay of pions and kaons into massless lepton pairs is
allowed.
 
The rate of energy loss due to one body decays is computed using the
standard formula \be \dot{\epsilon}_{\rm 1B}= \int
\frac{d^3p}{(2\pi)^32E} \int \frac{d^3k_1}{(2\pi)^32E_1}~E_1 \int
\frac{d^3k_2}{(2\pi)^32E_2} |A|^2 (2\pi)^4 \delta^4(P-k_1-k_2) \ee
where $P$, $k_1$ and $k_2$ are the four momenta of the meson, the
neutrino and the charged lepton respectively. We neglect the electron
mass in the describing the kinematics of these reaction because the
typical electron momentum is large. Further, since the electron
chemical potential we find is also small (compared to $T$) we are
justified in neglecting the final state electron Pauli-blocking. This
allows us to perform the integration over the final state lepton
momenta using the identity

\be 
\int \frac{d^3k_1}{2E_1} \int \frac{d^3k_2}{2E_2}\delta^4(P-k_1-k_2)
~k_1^\mu~k_2^\nu = \frac{\pi}{24} (P^\mu P^\nu+g^{\mu \nu} P^2)
\ee 

\ni to obtain

\be
 \dot{\epsilon}_{\rm 1B}= \frac{G_F^2~f_\pi^2~C^2~}{12 \pi} 
\int \frac{d^3p}{(2\pi)^3} ~f(E_p)~ ((\tilde{P}\cdot P)^2 - \tilde{P}^2~P^2)
\, .
\ee

\ni where $P=(E_p,\vec{p})$ is the four momentum of the Goldstone boson
and $C=\sin{\theta_C}$ for kaon decays. $C=\cos{\theta_C}$ for charged
pion decays and $C=1$ for the neutral current decay of the
$\pi^0$. Decay processes can occur only when the meson four momentum
is time like. Thus, the 1-body decay of $H$ boson is forbidden
and only low momentum pions and kaons (See Eq.~(\ref{space})) can
participate in the decay process. This accounts for the saturation of
the one body decay contribution to the emissivity with increasing
temperature seen in Fig. 4.

Higher momentum, space like, states can participate in processes such
as $e^{\pm} \phi^{\mp}\rightarrow \nu$.  We find that these lepton
absorption processes dominate the emissivity for $T \gsim 10 MeV$.
Processes involving two mesons in the initial state such as $\phi^\pm
\phi^0 \rightarrow e^\pm \nu$ and $\phi^+ \phi^- \rightarrow \nu
\bar{\nu}$ can be expected to become important only at high
temperature. This is because Goldstone bosons are weakly interacting,
the two particle amplitude is reduced by the factor $p/f_\pi$ compared
to the one body amplitude, where $p$ is the relative momenta of the
Goldstone bosons. Despite this suppression we consider the two meson
annihilation process because the pion and kaon densities exceeds the
lepton density at high temperature
\footnote{It is interesting to note that in vacuum, with normal
dispersion relations, the two body processes would dominate over one
body decay at modest temperatures due to angular momentum restrictions
on decay process.}.
\begin{figure}[t]
\begin{center}
\includegraphics[width=.8\textwidth]{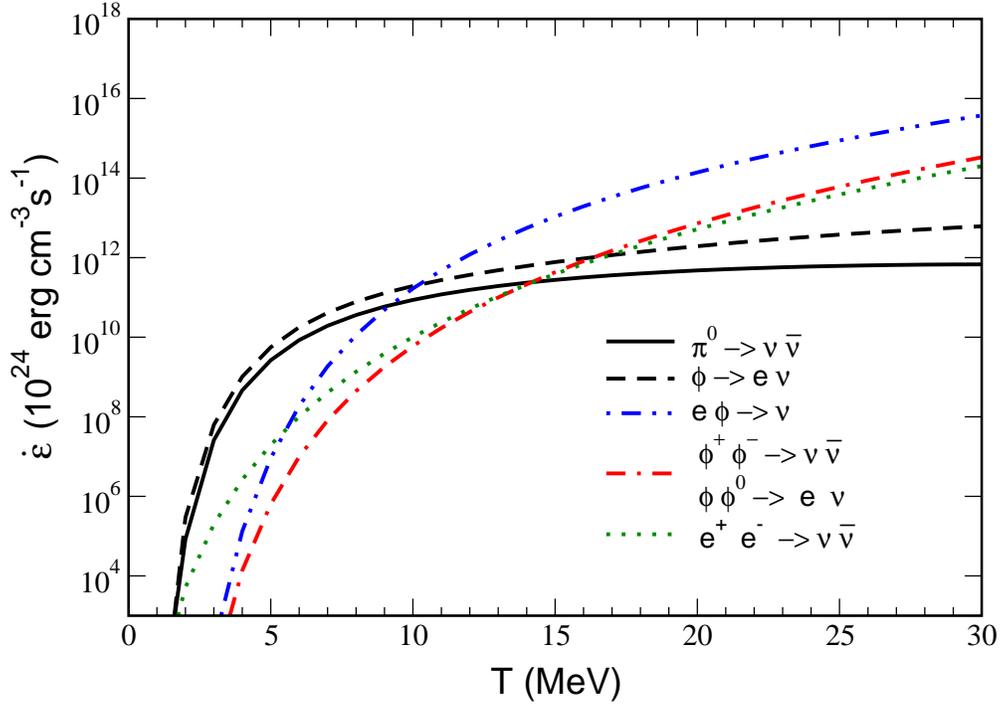}
\end{center}
\caption{Rate of energy loss due to neutrino emitting reactions. Contributions due to one body decay (solid curve), electron absorption on mesons (dot-dashed curve) and process involving the annihilation of two mesons is shown. The 
purely leptonic process $e^+ e^- \rightarrow \nu \bar{\nu}$ is also
shown (double dot-dashed curve).}
\label{emissivity}
\end{figure}

In Fig. 4, the rate of energy loss due to meson decay, lepton assisted
decays ($e \phi \rightarrow \nu$) and two meson reactions are
shown. For reasons discussed earlier, meson decay is the dominant
reaction for $T \lsim 10 $ MeV. Reactions involving two particles in
the initial state are also shown and become important at higher
temperature. Charged current absorption of electrons and positrons on
charged mesons producing neutrinos dominates the emissivity for
$T\gsim 10$ MeV. The two meson annihilation reactions contribute to
only 10\% of the total rate even at $T \sim 30$ MeV. The purely
leptonic process $ e^+ e^- \rightarrow \nu \bar{\nu}$ is also shown in
the figure. Its contribution is always smaller than the contribution
arising from the mesons.

Other processes involving electrons in the initial state, such as $e^-
\phi^+ \rightarrow \nu \phi^0$ could become important at low
temperature where the electron density exceeds the meson density. We
have ignored their contribution since this corresponds to $T\lsim 5$
MeV where the one-body decay dominates. We also note when the
temperature becomes comparable to the gap $\Delta$ neutrino emission
reactions involving quark quasi-particles can become important. In
particular, quark Cooper pair breaking has been shown to be relevant
when $T\sim \Delta$ \cite{Jaikumar:2001hq}. We expect the rate of this
process to small  at $T\sim 30$ MeV since it is suppressed by the
factor $\exp{(-2\Delta/T)}$.

\section{Conclusions}
We have shown that novel processes in which mesons are either emitted
or absorbed from neutrinos occur in the CFL plasma and contribute to
neutrino opacity. Absorption of, thermal, massless $H$ bosons is
the dominant reaction contributing to the neutrino opacity in the CFL
phase for temperatures in the range $T = 1-30$ MeV. Cherenkov
radiation of these mesons is the second most import process and is
likely to be the dominant process for $T \lsim 1$ MeV.  With
increasing temperature and the exponential growth of pion and kaon
number densities in the CFL plasma additional reactions such as $\nu +
\pi^0 \rightarrow \nu$ and $\nu_e + \pi^- \rightarrow e^-$ contribute
to the neutrino opacity.  We find that the mean free path for thermal
neutrinos at $T=10$ MeV is of the order of $10$ meters and at $T=5$
MeV it is similar to $100$ meters. In the table below we compare the
neutrino mean free path in CFL matter with those in nuclear matter and
unpaired quark matter under similar conditions. To make these
comparisons we employ earlier estimates of neutrino mean free path in
nuclear matter obtained by Reddy, Prakash and Lattimer
\cite{Reddy:1997yr} and in unpaired quark matter obtained by Iwamoto
\cite{iwamoto}. The quark chemical potential $\mu=400$ MeV corresponds
to a baryon density $n_B \sim 5 n_0$ in quark matter, where $n_0=0.16$
fm$^{-3}$ is the nuclear saturation density. In the table below, we
compare mean free path of thermal neutrinos ($E_\nu = \pi T$) in these
different phases at $n_B=5 n_0$.
\begin{center}
\begin{tabular}{|c|c|c|c|} \hline 
phase & process & $\lambda$(T=5 MeV) &$\lambda$(T=30 MeV) \\  \hline
Nuclear & $\nu n \rightarrow \nu n $ &  200 m & 1 cm \\ \cline{2-4}
Matter  & $\nu_e n \rightarrow e^- p $ &  2 m & 4 cm \\ \hline
Unpaired & $\nu q \rightarrow \nu q$ & 350 m & 1.6 m \\ \cline{2-4}
Quarks  &  $\nu d \rightarrow e^- u$ & 120 m & 4 m \\ \hline
CFL & $\lambda_{3B} $ & 100 m & 70 cm \\ \cline{2-4}
    & $\nu \phi \rightarrow \nu \phi$ & $>$10 km & 4 m \\ \hline
\end{tabular}
\end{center}
The findings presented in the table above indicate that neutrino mean
free path in CFL matter is similar to or shorter than that in unpaired
quark matter in the temperature range $T=1-30$ MeV. This, surprising,
result arises solely due to the novel processes involving Cherenkov
absorption and radiation of CFL Goldstone bosons.

Novel neutrino emitting processes such as $\pi^0$ decay to
neutrino-anti neutrino pairs occur in the CFL phase. Charged current
leptonic decay of mesons is also enhanced in the medium. They occur
because, in the rest frame of the meson moving relative to the medium,
the ground state breaks rotational invariance. With increasing
temperature, this enhancement saturates since only low momentum mesons
have a time like dispersion relation.  At higher temperatures,
reactions involving two particles overcome this kinematic constraint
and dominate the emissivity. In the neutron star context, the high
temperature emissivity is unlikely to play an important role in
neutron star dynamics because neutrinos are effectively trapped and
are described by local thermal distributions. We will restrict
ourselves to $T \lsim 10$ MeV to make the comparisons between the
emissivity of the CFL phase with that of unpaired quark matter. In
unpaired quark matter, Iwamoto finds that the emissivity due
$\beta$eta decay of light quarks at $n_B=5 n_0$ to be
$\dot{\epsilon}_{q\beta} \sim 2 \times 10^{36}$ erg cm$^{-3}$ s$^{-1}$
at $T=5$ MeV and $\dot{\epsilon}_{q\beta} \sim 2 \times 10^{38}$ erg
cm$^{-3}$ s$^{-1}$ at $T=10$ MeV \cite{iwamoto}. This is to be
compared with our finding that $\dot{\epsilon}_{CFL} \sim 5 \times
10^{33}$ erg cm$^{-3}$ s$^{-1}$ at $T=5$ MeV and $\dot{\epsilon}_{CFL}
\sim 2 \times 10^{35}$ erg cm$^{-3}$ s$^{-1}$ at $T=10$ MeV. The
emissivity is roughly reduced by three orders of magnitude. At lower
temperature, emissivity in the CFL phase is exponentially suppressed,
due to the paucity of thermal mesons, by the factor $\exp{(-m/T)}$,
where $m$ is the mass of the lightest octet meson. For a detailed
discussion of neutrino emission in the CFL phase at sub MeV
temperature, we refer the reader to an article by Jaikumar, Prakash
and Schafer which was posted on the archive concurrently with this
work \cite{Jaikumar:2002vg}.

The significant new finding of this work is that the neutrino opacity of
the CFL phase is similar or greater than that of unpaired quark
matter. Relative to nuclear matter the CFL phase is only marginally
less opaque. At lower temperature, it is roughly comparable to the
opacity nuclear matter given that the latter is known only to within
factor of a few at the densities or relevance to neutron stars.
Despite the energy gap in the quark quasi-particle excitations
spectrum the opacity remains large and there is {\it no} exponential
suppression of neutrino cross sections even when $T \ll \Delta$. This
is in sharp contrast to earlier findings of exponentially suppressed
neutrino cross sections in the two flavor superconducting phase of
quark matter which is devoid of Goldstone bosons in its excitation
spectrum \cite{Carter:2000xf}. The astrophysical implication of this
finding is that temporal aspects of neutrino diffusion inside the
newly born neutron star could be similar to that found in earlier
studies of neutrino transport in unpaired quark
matter\cite{Pons:2001ar}. However, since we expect the specific heat
of the CFL phase $C_V \sim T^3$ to be small compared to that of
unpaired quark matter where $C_V \sim \mu^2 T$ the cooling rates could
still differ and needs to be investigated. If the rate of energy loss
via neutrino diffusion depends solely on neutrino mean free path, 
we can expect the smaller heat capacity to result in accelerated
cooling due to a relatively more rapid rate of change of the ambient
temperature.  Nonetheless, we emphasize that the early evolution of a
newly born neutron star is a complex process which depends on several
microscopic ingredients and macroscopic conditions. In order to gauge
how color superconductivity in the neutron star core will impact
observable aspects of early neutron star evolution the rates computed
in this work, and in addition, the thermodynamic properties of the CFL phase
need to be included in detailed numerical simulations of core collapse
supernova. This is the only reliable means to bridge the gap between
the exciting theoretical expectation of color superconductivity at
high density and observable aspects of core collapse supernova --
the neutrino count rate, the neutrino spectrum and the explosion
itself.


\vskip0.75in \centerline{\bf Acknowledgments} We thank David Kaplan,
Krishna Rajagopal, Misha Stephanov and Frank Wilczek for several
useful discussions. We thank Thomas Schafer for spotting an error in
an early version of our manuscript and for discussions relating to 
the coupling of $U(1)_B$ Goldstone boson to the weak neutral
current. This work is supported in part by funds provided by the
U.S. Department of Energy (D.O.E.) under cooperative research
agreement DF-FC02-94ER40818. M. S. was supported by a fellowship from
the Foundation for Polish Science.  M. S. was also supported in part
by the Polish State Committee for Scientific Research, (KBN) grant
no. 2P 03B 094 19. M. T. was supported in part by Grant-in-Aid for
Scientific Research from Ministry of Education, Science, Sports and
Culture of Japan (No. 3666).

\end{document}